\documentclass[11pt,a4paper]{article}
\usepackage{times,latexsym}
\usepackage{url}
\usepackage[T1]{fontenc}

    \usepackage[acceptedWithA]{tacl2018v2}
\usepackage{xspace,mfirstuc,tabulary}

\usepackage{multirow}
\usepackage{algorithm}
\usepackage{algorithmic}
\usepackage{graphicx}
\newcommand{\partAdata}{\mathcal A}
\newcommand{\partBdata}{\mathcal B}
\newcommand{\atitle}{\mathcal T}
\newcommand{\acontent}{\mathcal C}

\newcommand{\satitlebatch}{\widetilde{t}}
\newcommand{\sacontentbatch}{\widetilde{c}}

\newif\iftaclinstructions
\taclinstructionsfalse 
\iftaclinstructions
\renewcommand{\confidential}{}
\renewcommand{\anonsubtext}{(No author info supplied here, for consistency with
TACL-submission anonymization requirements)}
\newcommand{\instr}
\fi

\iftaclpubformat 

\else

\fi

\title{Federated Hierarchical Hybrid Networks for Clickbait Detection}

\author{
 Feng Liao$^1$ \and Hankz Hankui Zhuo$^1$ \and Xiaoling Huang$^1$ \and Yu Zhang$^2$ \\
 $^1$Sun Yat-Sen University, Guangzhou, China; $^2$Arizona State University, US \\
  $^1${\sf \{liaof3@mail2,zhuohank@mail,huangxl29@mail2\}.sysu.edu.cn} \\
  $^2${\sf yu.zhang.442@asu.edu} \\
}

\date{}

\begin{document}
\maketitle
\begin{abstract}
	Online media outlets adopt clickbait techniques to lure readers to click on articles in a bid to expand their reach and subsequently increase revenue through ad monetization. As the adverse effects of clickbait attract more and more attention, researchers have started to explore machine learning techniques to automatically detect clickbaits. Previous work on clickbait detection assumes that all the training data is available locally during training. In many real-world applications, however, training data is generally distributedly stored by different parties (e.g., different parties maintain data with different feature spaces), and the parties cannot share their data with each other due to data privacy issues. It is challenging to build models of high-quality federally for detecting clickbaits effectively without data sharing. In this paper, we propose a federated training framework, which is called federated hierarchical hybrid networks, to build clickbait detection models, where the titles and contents are stored by different parties, whose relationships must be exploited for clickbait detection. We empirically demonstrate that our approach is effective by comparing our approach to the state-of-the-art approaches using datasets from social media.
\end{abstract}

\section{Introduction}
Clickbait is a text or thumbnail link that is designed to entice users to access the linked online content, which often fails to fulfill the promise made by the title. Online media outlets adopt clickbait techniques in a bid to expand their reach and subsequently increase revenue through ad monetization. As the adverse effects of clickbait attract more and more attention, researchers have started to explore machine learning techniques to automatically detect clickbaits. For example, previous approaches, such as \cite{DBLP:conf/asunam/ChakrabortyPKG16,DBLP:journals/pacmhci/ChakrabortySMG17,DBLP:conf/ijcai/WeiW17}, built text classifiers via feature engineering methods such as \cite{DBLP:conf/asunam/RonyHY17,DBLP:journals/symmetry/ZhengCYSJZ18,DBLP:conf/ecir/Anand0P17,DBLP:journals/corr/abs-1710-05364,DBLP:conf/sigir/KumarKGLV18} built text classifiers using deep learning models. Despite the success of previous approaches, they all assume that all the training data is stored locally and hence the detection models can be built locally. 

However, in reality, different organizations generally hold different parts of the data and cannot share data with each other \cite{DBLP:journals/tist/YangLCT19}. \emph{For example, a social network company, such as Twitter, Weibo, etc., often have the need to automatically monitor the quality of content via a clickbait detection model. While they may access a title such as "LeBron James was dragged along the street", they do not have access to its corresponding externally linked content, such as "A man walked down the street with LeBron James' autobiography in his hand. The cover of the autobiography is a picture of James."}

In the above scenario, we found that a qualified model is difficult to obtain using prior work since the local data alone is not enough for clickbait detection. In our experiments, we found that the correlation between titles and contents is a key factor for clickbait detection while most of the prior work failed to exploit the correlation.

In response to the above problem, we propose Federated Hierarchical Hybrid Networks, which is Hierarchical Hybrid Networks trained by Clickbait Federated Learning. Hierarchical Hybrid Networks exploit the connection between title and content, which is a key factor for clickbait detection. Clickbait Federated Learning can effectively utilize data from two parties for model training without the agreement of the network structures from the two parties. Our experimental results show that Federated Hierarchical Hybrid Networks outperform other clickbait detection models, and is comparable to the models trained in the ideal situation.

We organize the paper as follows. We first review related work. After that, we present the details of our framework and then give a detailed description of our algorithm. Finally, we evaluate our algorithm in a dataset and conclude our work with a discussion on future work.

\section{Related Work}
\subsection{Clickbait Detection}
As we mentioned, there has been a substantial amount of prior work on automatic clickbait detection. Chakraborty et al. collected extensive titles for both clickbait and non-clickbait categories and manually extracted features for SVM classifier, Decision Trees classifier, and Random Forests classifier \cite{DBLP:conf/asunam/ChakrabortyPKG16}. As an extension, Chakraborty et al. used another dataset collected from Twitter to conduct further analysis of clickbait \cite{DBLP:journals/pacmhci/ChakrabortySMG17}. In the work of Wei and Wan, they redefined the clickbait problem and identified ambiguous and misleading titles separately. Wei and Wan crawled a total of 40000 articles from four major Chinese news sites and used SVM classifier for text classification with manual feature extraction \cite{DBLP:conf/ijcai/WeiW17}.

FastText \cite{DBLP:conf/eacl/GraveMJB17}, TextCNN \cite{DBLP:conf/emnlp/Kim14}, TextRNN \cite{DBLP:conf/emnlp/ChoMGBBSB14}, and Self-Attentive Network \cite{DBLP:conf/naacl/YangYDHSH16,DBLP:journals/corr/LinFSYXZB17} are classic deep learning models. Rony et al. applied a model similar to FastText, which uses distributed sub-word embedding learned from a large corpus to detect clickbait \cite{DBLP:conf/asunam/RonyHY17}. Zheng et al. proposed CBCNN. The architecture of CBCNN is similar to TextCNN \cite{DBLP:journals/symmetry/ZhengCYSJZ18}. In the work of Anand et al., a model similar to TextRNN was applied to clickbait detection. Their model combines Distributed Word Embedding with Character Level Word Embedding \cite{DBLP:conf/ecir/Anand0P17}. The first attempt to apply Self-Attentive Network to clickbait detection is in the work of Zhou \cite{DBLP:journals/corr/abs-1710-05364}.

The connection between title and content is an important feature in the clickbait detection task. To the best of our knowledge, only one prior work takes advantage of the connection \cite{DBLP:conf/sigir/KumarKGLV18}. In the work of Kumar et al., they utilized not only the similarity between title and description but also the similarity between description and image. Their work emphasizes the similarity between different parts of the article while we focus on whether the title is ambiguous or misleading (Marquez divided news headlines into three types: accurate, ambiguous and misleading \cite{Marquez1980How}), which is an important indicator of clickbait. Notably, due to the flexibility of natural language, similarity cannot accurately measure the connection between title and content for clickbait detection. 

\subsection{Federated Learning}
The Data Island problem has attracted more and more attention. Mcmahan et al. and Konen et al. advocated an alternative that left the training data distributed on the mobile devices and learned a shared model by aggregating locally computed updates \cite{DBLP:journals/corr/McMahanMRA16,DBLP:journals/corr/KonecnyMRR16}. Konen et al. proposed two ways to reduce the uplink communication costs, which improved the efficiency of their framework of federated learning \cite{DBLP:journals/corr/KonecnyMYRSB16}. Smith et al. proposed a novel systems-aware optimization framework for federated multi-task learning and this method achieved significant speedups compared to alternatives in the federated setting \cite{DBLP:conf/nips/SmithCST17}. Yang et al. introduced a comprehensive secure federated-learning framework, which includes horizontal federated learning, vertical federated learning, and federated transfer learning \cite{DBLP:journals/tist/YangLCT19}. Zhuo et al. propose federated reinforcement learning, which considers the privacy requirement and builds Q-network for each agent with the help of other agents \cite{DBLP:journals/corr/abs-1901-08277}.

\section{Problem Definition}
Our problem can be defined by: given as input a set of pairs of titles and labels $Data_{A}=(\partAdata, \gamma)$ from $PartyA$, and a set of contents $Data_{B}=(\partBdata, \gamma)$ from $PartyB$, find the model $\delta$ that must be built on both $Data_{A}$ and $Data_{B}$ while $PartyA$ cannot share data with $PartyB$. This problem is called the Data Island problem.

According to the above setting, in a clickbait detection task, $PartyA$ is the social network company, which needs to monitor the quality of content on its platform. And $PartyB$ is the media company. $t$ is a link to an external article, which contains the textual description of this article. $t$ is composed of a sequence of words $t=\{w_1,\cdots,w_i\}$ and it can be regarded as the title of the article. $l$ is a label in $\{clickbait, non\textrm{-}clickbait\}$, specifying whether the corresponding title $t$ is a clickbait or not. We denote the set of all pairs of titles and labels as $\partAdata$ ($\partAdata = \{\langle t, l \rangle \}$). $c$ is the detailed content from $PartyB$ corresponding to title $t$ from $PartyA$, which is composed of a sequence of words $c=\{w_1,\cdots,w_j\}$. We denote the set of all contents as $\partBdata$ ($\partBdata = \{\langle c \rangle \}$). An article $a$ is made up of title $t$ and content $c$. By mapping $\gamma$, we can find the title $t$ and the content $c$ from the same article $a$. $\delta(a)=l$ represents the automatic clickbait detection process.

In a word, $PartyA$ hosts $\partAdata$, and $PartyB$ hosts $\partBdata$. They cannot share data with each other. It meaning that $\partAdata$ and $\partBdata$ cannot be aggregated for model training.

\section{Our Approach}
Inspired by human practices in clickbait detection, we propose Hierarchical Hybrid Networks, which exploits the connection between title and content, while most of the previous approaches either fail to consider both or only take their similarity into account. However, like all of the prior works, Hierarchical Hybrid Networks only work in the ideal situation (data sharing). So we propose here a novel method of training model: Clickbait Federated Learning as an effective solution to the Data Island problem. This method can effectively utilize data from two parties for model training. What is more, Clickbait Federated Learning does not require the two parties to agree on the network structures. After we train Hierarchical Hybrid Networks by Clickbait Federated Learning, we obtain Federated Hierarchical Hybrid Networks, which provides a solution to the Data Island problem.

\subsection{Hierarchical Hybrid Networks}
\begin{figure}[!ht]
	\centering
	\includegraphics[width=0.4\textwidth]{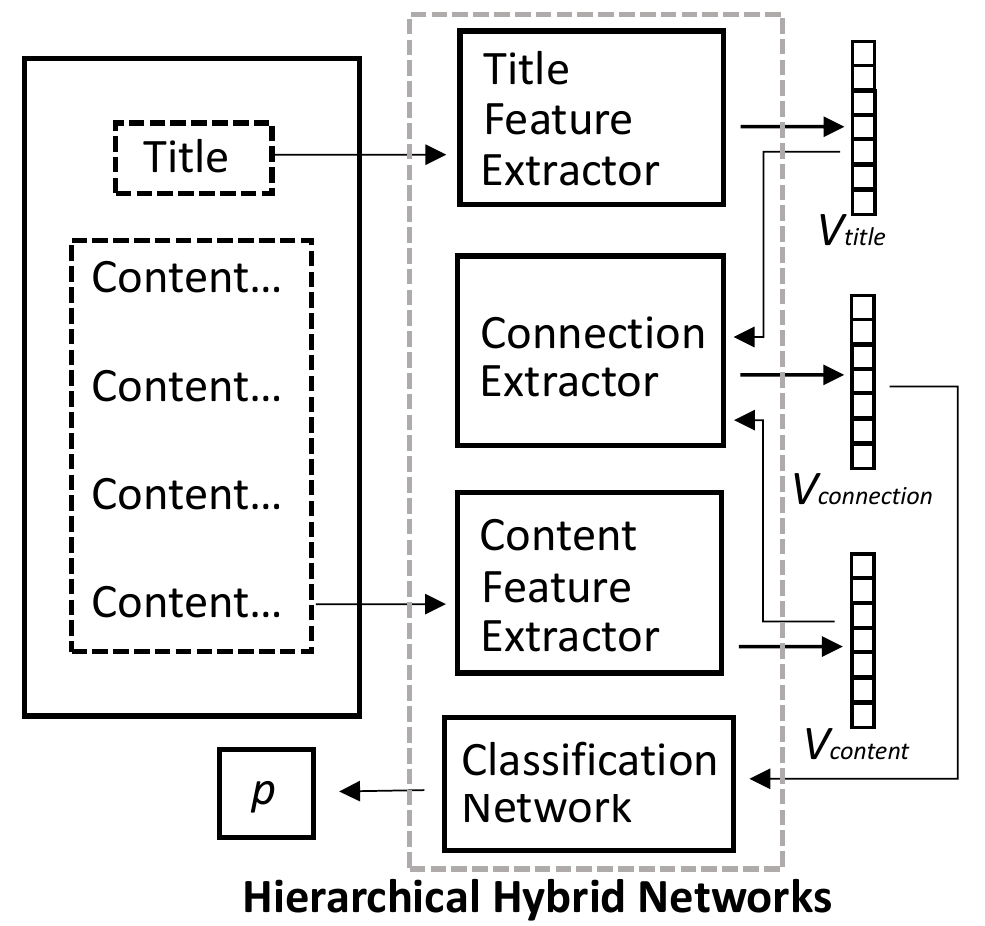}
	\caption{The architecture of Hierarchical Hybrid Networks.}
	\label{architecture}
\end{figure}

As shown in Figure \ref{architecture}, Hierarchical Hybrid Networks consists of four parts: title feature extractor, content feature extractor, connection extractor, and classification network. The hypothesis of Hierarchical Hybrid Networks is in the ideal situation (data sharing).

Before the modeling, we applied the following preprocessing procedure: remove illegal characters and stop words, word tokenize, etc. After this process, we get a standardized title and content of an article. We denote a batch of standardized titles as $\satitlebatch$, a batch of standardized contents as $\sacontentbatch$.

And then we feed $\satitlebatch$ and $\sacontentbatch$ as inputs to Hierarchical Hybrid Networks to get prediction $p$. We denote $\mathcal{M}_1$, $\mathcal{M}_2$, $\mathcal{M}_3$, $\mathcal{M}_4$ as title feature extractor, content feature extractor, connection extractor, classification network respectively. And $\theta_1$, $\theta_2$, $\theta_3$, $\theta_4$ are the parameters of $\mathcal{M}_1$, $\mathcal{M}_2$, $\mathcal{M}_3$, $\mathcal{M}_4$ respectively. After feeding $\satitlebatch$ into $\mathcal{M}_1$, we obtain the title feature vector $V_{title}$. In a similar way, we obtain the content feature vector $V_{content}$. We concatenate $V_{title}$ and $V_{content}$ and feed it to $\mathcal{M}_3$ to obtain the connection vector $V_{connection}$. Based on $V_{connection}$, the $\mathcal{M}_4$ makes a prediction. With label $l$, we calculate the loss and update the parameters in the training process. In the predicting process, we get the corresponding label based on $p$. The above process is shown in Equation \ref{eq1}:
\begin{equation}\label{eq1}
\begin{array}{l}
V_{title} = \mathcal{M}_1(\satitlebatch;\theta_1) \\
V_{content} = \mathcal{M}_2(\sacontentbatch;\theta_2) \\
V_{connection} = \mathcal{M}_3(V_{title}, V_{content};\theta_3) \\
p = \mathcal{M}_4(V_{connection};\theta_4)
\end{array}
\end{equation}

\subsubsection{Feature Extractor}
\begin{figure}[!ht]
	\centering
	\includegraphics[width=0.5\textwidth]{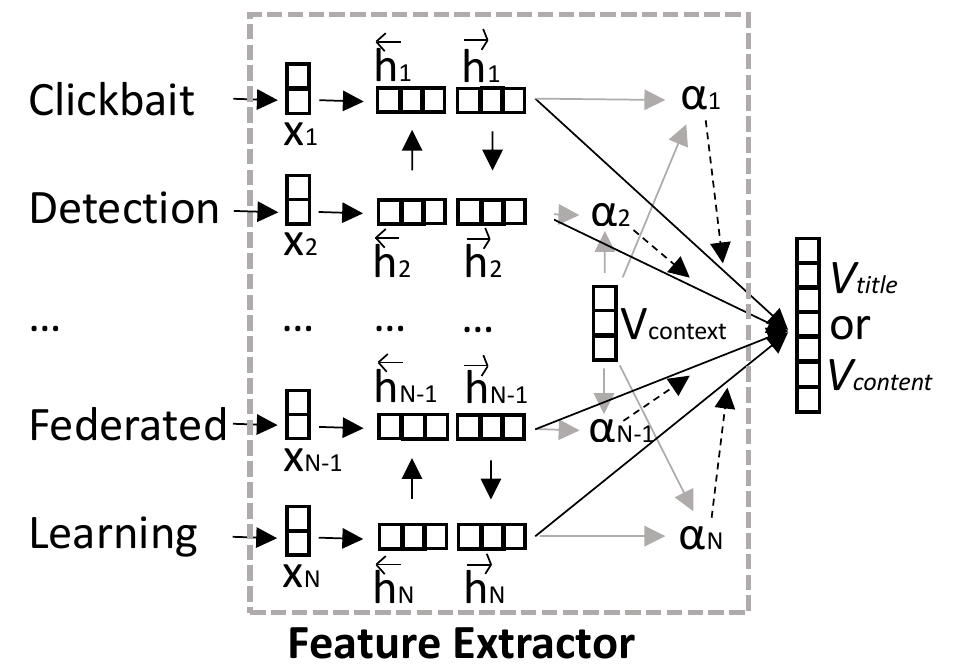}
	\caption{The implementation of the title feature extractor and content feature extractor.}
	\label{FeatureExtractor}
\end{figure}

Since clickbait detection is a text classification task, we need a feature extractor to extract features from the text for classification. We applied Self-Attentive Network to implement $\mathcal{M}_1$ and $\mathcal{M}_2$ as shown in Figure \ref{FeatureExtractor}. Given a title~(or content) that contains N tokens, we first map each token $w_i$, where i $\in$ [1, N], to its corresponding word embedding $\mathrm{x_{i}}$, through a word embedding matrix~(100-dimension pre-trained Glove embedding of Wikipedia data \cite{DBLP:conf/emnlp/PenningtonSM14}).

After that, we use a bi-directional LSTM \cite{DBLP:journals/neco/HochreiterS97} to encode the contextual information from both directions of the token into its hidden state. The resulting hidden state of BiLSTM for each token was the concatenation of its forward hidden state and backward hidden states, as shown in the following equation:
\begin{equation}\label{eq21}
\begin{array}{l}
\overrightarrow{\mathrm{h_{i}}} = \overrightarrow{LSTM}(\mathrm{x_{i}}, \overrightarrow{\mathrm{h_{i-1}}}) \\
\overleftarrow{\mathrm{h_{i}}} = \overleftarrow{LSTM}(\mathrm{x_{i}}, \overleftarrow{\mathrm{h_{i+1}}}) \\
\mathrm{h_{i}} = \overrightarrow{\mathrm{h_{i}}} \parallel \overleftarrow{\mathrm{h_{i}}}
\end{array}
\end{equation}

We concatenate all hidden state and get $\mathbf{H}$. The token level attention vector $\alpha$ represents the weights of tokens. $\mathbf{W_{a}}$ and $\mathbf{V_{context}}$ is the network and the context vector of attention mechanism. Both of them are the parameters to train. The process is shown in Equation \ref{eq22}:
\begin{equation}\label{eq22}
\begin{array}{l}
\mathbf{H} = [\mathrm{h_{1}},\cdots,\mathrm{h_{N}}] \\
\alpha = softmax(tanh(\mathbf{H}\mathbf{W_{a}})\mathbf{V_{context}}) \\
V_{title} = \mathbf{H}^\mathrm{T}\alpha,\quad\quad\, if\ input\ is\ \satitlebatch \\
V_{content} = \mathbf{H}^\mathrm{T}\alpha,\quad if\ input\ is\ \sacontentbatch
\end{array}
\end{equation}

\subsubsection{Connection Extractor}
\begin{figure}[!ht]
	\centering
	\includegraphics[width=0.5\textwidth]{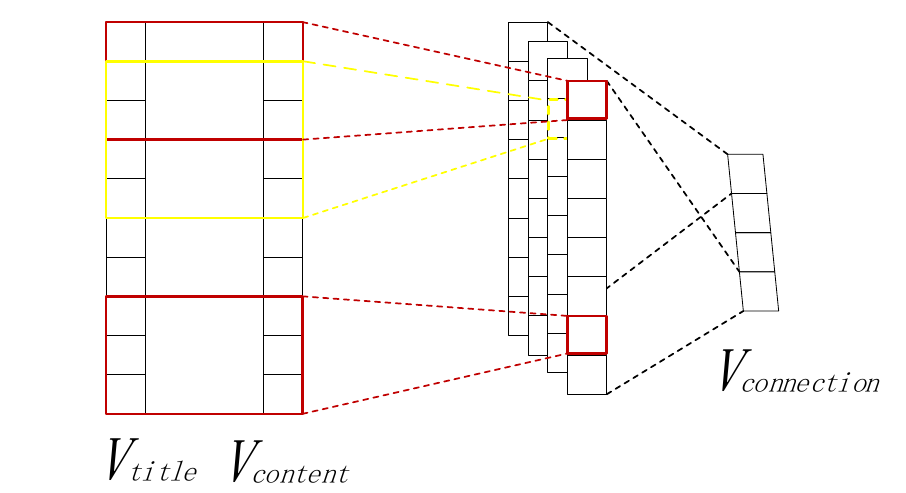}
	\caption{The implementation of the connection extractor.}
	\label{ConnectionExtractor}
\end{figure}

The clickbait problem stems from the content of the article failing to fulfill the promise made by the title of the article. So when detecting clickbait, human combines the title with the content. Based on the complex connection between title and content, human makes the final judgment. Inspired by human practices in clickbait detection, we design a connection extractor implemented by Convolutional Neural Network, as shown in Figure \ref{ConnectionExtractor}.

After we obtain the vector $V_{title}$ and $V_{content}$ from feature extractors, we concatenate them and get $V$. We apply a convolution operation to get the connection feature. $\mathbf{W_{c}}$ and $\mathbf{b_{c}}$ are the parameters of this filter and $f$ is the activation function. A feature $c_{i}$ is learned from the $i$th row to the $i+h-1$th row of $V$. The feature map $c$ is the concatenation of all features. The process is as Equation \ref{eq11}:
\begin{equation}\label{eq11}
\begin{array}{l}
V = V_{title} \parallel V_{content} \\
c_{i} = f(\mathbf{W_{c}}V_{i:i+h-1} + \mathbf{b_{c}}) \\
c = [c_{1},\cdots,c_{n-h+1}]
\end{array}
\end{equation}

We then apply a max pooling operation over the feature map $c$ to obtain $\hat{c_{h}}$ as the final feature corresponding to this particular filter with height $h$. The concatenation of all final feature $\hat{c_{h}}$ is the connection vector $V_{connection}$ we need, as shown below:
\begin{equation}\label{eq12}
\begin{array}{l}
\hat{c_{h}} = max\{c\} \\
V_{connection} = \hat{c_{h}} \parallel \cdots \parallel \hat{c_{k}}
\end{array}
\end{equation}

By this connection extractor, $V_{connection}$ contains not only the connection between title and content but also the features of title and content.

\subsubsection{Classification Network}
$\mathcal{M}_4$ is a fully connected neural network. $\mathbf{W}$ and $\mathbf{b}$ is the parameters of the classification network. The process is given by the following equation:
\begin{equation}\label{eq5}
\begin{array}{l}
p = softmax(\mathbf{W}V_{connection}+\mathbf{b})
\end{array}
\end{equation}

\subsection{Clickbait Federated Learning}
\begin{algorithm}[!ht]
	\caption{Clickbait Federated Learning in $PartyA$}\label{algo1}
	\textbf{Input:} $\partAdata = \{\langle t, l \rangle \}$\\
	\textbf{Output:} $\mathcal{M}_1$ and $\mathcal{M}_4$ of federated model $M_{FED}$
	
	\begin{algorithmic}[1]
		\STATE identify the overlapping data in $PartyA$;
		\STATE determine every batch and synchronize it with $PartyB$;
		\STATE preprocess title $t$ get $\satitlebatch$;
		\STATE initialize $\mathcal{M}_1$ and $\mathcal{M}_4$;
		\STATE input corresponding batch $\satitlebatch$ to $\mathcal{M}_1$ get $V_{title}$;
		\STATE wait for $PartyB$ send $V_{content}$;
		\STATE concatenate $V_{title}$ and $V_{content}$, and then input it to $\mathcal{M}_4$, get prediction $p$ and loss;
		\STATE calculate $\frac{\mathrm{d}loss}{\mathrm{d}\theta_1}$, $\frac{\mathrm{d}loss}{\mathrm{d}\theta_4}$ and $\frac{\mathrm{d}loss}{\mathrm{d}V_{content}}$;
		\STATE update $\mathcal{M}_1$ and $\mathcal{M}_4$ with $\frac{\mathrm{d}loss}{\mathrm{d}\theta_1}$ and $\frac{\mathrm{d}loss}{\mathrm{d}\theta_4}$;
		\STATE send $\frac{\mathrm{d}loss}{\mathrm{d}V_{content}}$ to $PartyB$;
		\STATE repeat above Step 5 to Step 10 until $M_{FED}$ converges;
		\RETURN $\mathcal{M}_1$ and $\mathcal{M}_4$;
	\end{algorithmic}
\end{algorithm}

The hypothesis of Hierarchical Hybrid Networks is in the ideal situation: aggregating the data together and training the model. But in the Data Island problem setting, we cannot aggregate the data together to train Hierarchical Hybrid Networks. Prior work on federated learning is to train a shared model with data scattered in a large number of nodes. The problem we focus on is to train a model with data stored in two companies. So we propose a novel model training method: Clickbait Federated Learning. This method can effectively utilize data from two parties for model training. Clickbait Federated Learning does not require the two parties to agree on the network structures. As a result, This method is a convenient and general method. 

According to the assumptions in Problem Definition, $PartyA$ has title $t$ and label $l$ while $PartyB$ has content $c$~(both of them can access to the labels since labels are generated for training model so labels can be exchanged). $PartyA$ cannot share data with $PartyB$. Our target is to find a way to train model in the Data Island setting.

As shown in Algorithm \ref{algo1} and Algorithm \ref{algo2}, Clickbait Federated Learning consists of two parts. In reality, the data stored in different places usually does not completely overlap. So we need to identify the overlapping data for training and give it a unique id. Due to the case of non-shared data, we use encryption~(Step 1). And then we determine every batch of every epoch and synchronize it on both sides so that we make sure the same samples are used in every batch of every epoch~(Step 2). This requires the coordination of both sides.

\begin{algorithm}[!ht]
	\caption{Clickbait Federated Learning in $PartyB$}\label{algo2}
	\textbf{Input:} $\partBdata = \{\langle c \rangle \}$~(content)\\
	\textbf{Output:} $\mathcal{M}_2$ of federated model $M_{FED}$
	
	\begin{algorithmic}[1]
		\STATE identify the overlapping data in $PartyB$;
		\STATE determine every batch and synchronize it with $PartyA$;
		\STATE preprocess content $c$ get $\sacontentbatch$;
		\STATE initialize $\mathcal{M}_2$;
		\STATE input corresponding batch $\sacontentbatch$ to $\mathcal{M}_2$ get $V_{content}$;
		\STATE send $V_{content}$ to $PartyA$;
		\STATE wait for $PartyA$ send $\frac{\mathrm{d}loss}{\mathrm{d}V_{content}}$;
		\STATE calculate $\frac{\mathrm{d}V_{content}}{\mathrm{d}\theta_2}$;
		\STATE update $\mathcal{M}_2$ with $\frac{\mathrm{d}loss}{\mathrm{d}V_{content}} \cdot \frac{\mathrm{d}V_{content}}{\mathrm{d}\theta_2}$;
		\STATE repeat above Step 5 to Step 9 until $M_{FED}$ converges;
		\RETURN $\mathcal{M}_2$;
	\end{algorithmic}
\end{algorithm}

Secondly, we apply the training data to the preprocessing procedure and obtain standardized title and standardized content respectively~(Step 3). Then we initialize feature extractor $\mathcal{M}_1$ and $\mathcal{M}_2$ on both sides~(Step 4). The actual implementation of feature extractor is not critical. $PartyA$ can apply Convolutional Neural Network or Recursive Neural Network to implement feature extractor and the same with $PartyB$. Neither $PartyA$ nor $PartyB$ know the specific implementation of each other. This ensures data privacy and security. After that, we initialize the classification network $\mathcal{M}_4$ for the side that has labels. Up to now, the preparatory work before training is done.

Next is the repetitive training steps. As shown in Algorithm \ref{algo1} and Algorithm \ref{algo2}, the training steps is different between $PartyA$ and $PartyB$. According to chronological order, after we input the corresponding batch $\satitlebatch$ and $\sacontentbatch$ to $\mathcal{M}_1$ and $\mathcal{M}_2$ get $V_{title}$ and $V_{content}$ respectively for $PartyA$ and $PartyB$~(Step 5), $PartyB$ sends $V_{content}$ to $PartyA$~(Step 6 in $PartyB$). When $PartyA$ receives $V_{content}$, $PartyA$ concatenates $V_{title}$ and $V_{content}$ then feeds it to $\mathcal{M}_4$~(Step 6, 7in $PartyA$). Based on the prediction $p$ and label, $PartyA$ calculates the relevant derivatives and update the parameters of $\mathcal{M}_1$ and $\mathcal{M}_4$~(Step 8, 9 in $PartyA$). After that, $PartyA$ sends $\frac{\mathrm{d}loss}{\mathrm{d}V_{content}}$ to $PartyB$~(Step 10 in $PartyA$).

When $PartyB$ receives $\frac{\mathrm{d}loss}{\mathrm{d}V_{content}}$~(Step 7 in $PartyB$), $PartyB$ updates the parameters of $\mathcal{M}_2$ with the dot product of $\frac{\mathrm{d}loss}{\mathrm{d}V_{content}}$ and $\frac{\mathrm{d}V_{content}}{\mathrm{d}\theta_2}$~(Step 8, 9 in $PartyB$). Repeat these above training steps until the federated model $M_{FED}$, which is composed of $\mathcal{M}_1$, $\mathcal{M}_2$ and $\mathcal{M}_4$, converges and then we have obtained the clickbait detection model.

As described above, part of the training sequence is critical in Clickbait Federated Learning. The Step 6 in $PartyB$ must precede the Step 6 in $PartyA$ and the Step 10 in $PartyA$ must precede the Step 7 in $PartyB$. If $M_{FED}$ converges, $PartyA$ will send a termination signal to $PartyB$. Hence, Clickbait Federated Learning requires the coordination of both sides while maintaining data privacy.

According to Chain Rule, the above dot product equals $\frac{\mathrm{d}loss}{\mathrm{d}\theta_2}$, which is the derivatives of the ideal situation: aggregating the data together and training the model. So theoretically, the effect of $M_{FED}$ equals the effect of the model $M_{IDEAL}$ with the same architecture that is trained in the ideal situation.

We can also assume $PartyA$ has $c$ and $l$ while $PartyB$ have $t$. We still can get an excellent federated model by Clickbait Federated Learning. What is more, we can generalize $t$ and $c$ to other data types. Clickbait Federated Learning can be generalized to any similar multi-input classification tasks with non-shared data. Hence Clickbait Federated Learning is a general vertical federated learning method and represents a solution for the Data Island problem.

\subsection{Federated Hierarchical Hybrid Networks}
\begin{figure}[!ht]
	\centering
	\includegraphics[width=0.5\textwidth]{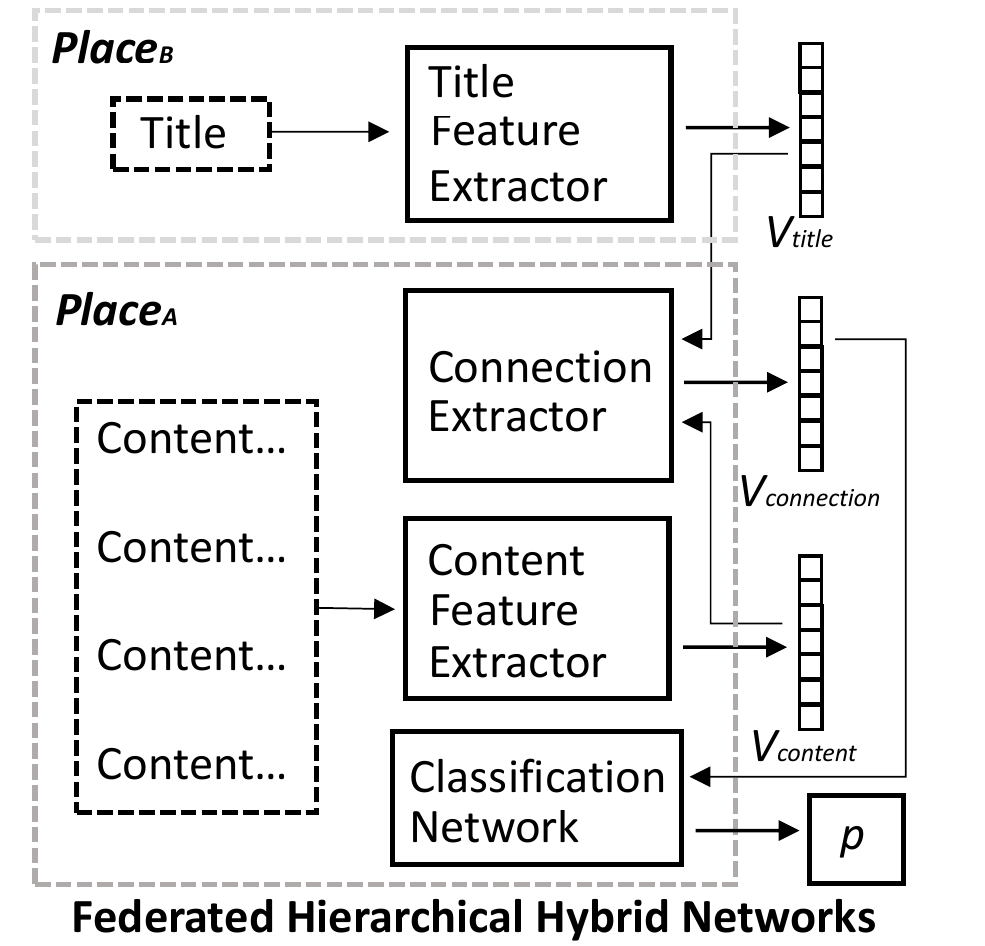}
	\caption{The architecture of Federated Hierarchical Hybrid Networks.}
	\label{FHHNarchitecture}
\end{figure}

In Clickbait Federated Learning, the connection extractor is not necessary since Clickbait Federated Learning just is a model training method. But if we append connection extractor with Convolutional Neural Network as the implementation in $PartyA$, and apply Self-Attentive Network to implement the feature extractor in $PartyA$ and the feature extractor in $PartyB$, we get Federated Hierarchical Hybrid Networks, whose architecture is the same as Hierarchical Hybrid Networks, as shown in Figure \ref{FHHNarchitecture}. Hence, Federated Hierarchical Hybrid Networks, which presents a solution to the Data Island problem, can be considered as Hierarchical Hybrid Networks trained by Clickbait Federated Learning.

\section{Experiments}
\subsection{Dataset}
We use the dataset provided by The Clickbait Challenge 2017~(http://www.clickbait-challenge.org/), which is a classic clickbait detection competition, in this paper. The provided dataset contains posts from a social media platform “Twitter”. This platform is often used by media to publish links to their websites. Each post, “tweet”, is a short message~(up to 140 characters), which can be accompanied by a link and a picture.

Each instance in the dataset includes $id$, $postText$~(the content of the tweet), $targetTitle$~(the title of the actual article), $targetParagraphs$~(the actual content of the article), $targetDescription$~(the description from the meta tags of the article), $targetKeywords$~(the keywords from the meta tags of the article), $targetCaptions$~(all captions in the article), $postMedia$~(the image that was posted alongside with the tweet), $truthClass$~(the clickbait label evaluated by five human evaluators), etc.

\begin{table}[!ht]
	\caption{Statistics of the datasets.}\label{tb1}
	\centering
	\begin{tabular}{|c|c|c|c|}
		\hline
		\textbf{Dataset} & tweets & clickbait & non-clickbait\\
		\hline
		A & 2459 & 762 & 1697\\
		\hline
		B & 19538 & 4761 & 14777\\
		\hline
	\end{tabular}
\end{table}

According to the source of the dataset, the above dataset is divided into two parts. The statistics of the two datasets is shown in Table \ref{tb1}. In our experiments, dataset A is the test set and dataset B is used as the training set and the validation set.

\subsection{Experiment Design}
As we mentioned, our problem is to design model $\delta$ between $a$ and $l$ in the situation of Data Island~(data not sharing). According to our approach, we design two experiments. One is to train Hierarchical Hybrid Networks and to compare the performance of it with other clickbait detection models in the ideal situation.

Another is to train models in the Data Island situation. In this situation, we only obtain $\partAdata = \{\langle t, l \rangle \}$ from $PartyA$. It means we can only train the model on titles and labels using the traditional training method. We also can train the model on contents and labels using the traditional training method. With Clickbait Federated Learning, we train a series of federated learning models with the same implementation of model architectures as these above models.

In our experiment, we choose $targetTitle$ as title and $targetDescription$ as content. Two evaluation metrics are used in this work: ROC-AUC and F1-score.

\subsection{Experimental Results}
To avoid the randomness effect, we perform all our experiments using 5-fold cross-validation on the above dataset. So the following experimental results reflect the average performance of each model on the test set. All of these models use 100-dimension pre-trained Glove embedding of Wikipedia data \cite{DBLP:conf/emnlp/PenningtonSM14}.

\subsubsection{The Ideal Situation}
In this setting, we can aggregate the title and content together for model training. We train five different models in this situation by the traditional training method. TextCNN~($\atitle$\&$\acontent$), TextRNN~($\atitle$\&$\acontent$), TextSAN~($\atitle$\&$\acontent$), and FastText~($\atitle$\&$\acontent$) have the same model architectures. All of them have a title feature extractor, a content feature extractor, and a classification network. The difference between them is the implementation of the title feature extractor and content feature extractor. 

\begin{table}[!ht]
	\caption{The experimental results in the ideal situation.}\label{tb2}
	\centering
	\begin{tabular}{|c|c|c|}
		\hline
		\multirow{2}{*}{\textbf{Model}} & \multicolumn{2}{c|}{\textbf{Measure}}\\
		\cline{2-3}
		& ROC-AUC & F1 \\
		\hline
		HHN & \textbf{0.67224} & \textbf{0.53155} \\
		\hline
		TextCNN($\atitle$\&$\acontent$) & 0.66185 & 0.51734 \\
		\hline
		TextRNN($\atitle$\&$\acontent$) & 0.65026 & 0.49115 \\
		\hline
		TextSAN($\atitle$\&$\acontent$) & 0.64851 & 0.48759 \\
		\hline
		FastText($\atitle$\&$\acontent$) & 0.64489 & 0.47466 \\
		\hline
	\end{tabular}
\end{table}

As shown in Table \ref{tb2}, Hierarchical Hybrid Networks have the best performance. We must give credit to the connection extractor since it is the only difference between Hierarchical Hybrid Networks and TextSAN~($\atitle$\&$\acontent$). We infer that the connection extractor extracts the complex connection between title and content effectively while retaining the original feature information in $V_{title}$ and $V_{content}$. This illustrates the importance of the correlation between title and content in the clickbait detection task.

\subsubsection{The Data Island}
In this setting, we cannot aggregate the title and content together for model training. So when using the traditional training method, we can only utilize the title or content for model training. ($\atitle$) means the model only accepts the title as input while ($\acontent$) means the model only accepts the content as input. In this situation, we train five federated learning models by Clickbait Federated Learning and eight models by the traditional training method.

Since $PartyA$ have $\partAdata = \{\langle t, l \rangle \}$, we train TextCNN~($\atitle$), TextRNN~($\atitle$), TextSAN~($\atitle$), and FastText~($\atitle$) independently. All of them have a title feature extractor and a classification network. The only difference between them is the implementation of feature extractor.

We also train TextCNN~($\acontent$), TextRNN~($\acontent$), TextSAN~($\acontent$), and FastText~($\acontent$) independently in $PartyB$. Similarly, all of them have a content feature extractor and a classification network. The only difference between them is the implementation of feature extractor.

\begin{table}[!ht]
	\caption{The experimental results in the Data Island.}\label{tb3}
	\centering
	\begin{tabular}{|c|c|c|}
		\hline
		\multirow{2}{*}{\textbf{Model}} & \multicolumn{2}{c|}{\textbf{Measure}}\\
		\cline{2-3}
		& ROC-AUC & F1\\
		\hline
		FedCNN & \textbf{0.65694} & \textbf{0.50881}\\
		\hline
		TextCNN($\atitle$) & 0.63943 & 0.47120\\
		\hline
		TextCNN($\acontent$) & 0.61256 & 0.44583\\
		\hline
		FedRNN & \textbf{0.64627} & \textbf{0.47744}\\
		\hline
		TextRNN($\atitle$) & 0.63306 & 0.47102\\
		\hline
		TextRNN($\acontent$) & 0.60779 & 0.42497\\
		\hline
		FedSAN & \textbf{0.64888} & \textbf{0.48603}\\
		\hline
		TextSAN($\atitle$) & 0.63555 & 0.46201\\
		\hline
		TextSAN($\acontent$) & 0.59554 & 0.40951\\
		\hline
		FedFastText & \textbf{0.64224} & \textbf{0.46843}\\
		\hline
		FastText($\atitle$) & 0.63287 & 0.45195\\
		\hline
		FastText($\acontent$) & 0.59350 & 0.39497\\
		\hline
		FedHHN & \textbf{0.66835} & \textbf{0.52898}\\
		\hline
	\end{tabular}
\end{table}

According to the experimental results in Table \ref{tb3}, title is more valuable for clickbait detection task than content since TextCNN~($\atitle$), TextRNN~($\atitle$), TextSAN~($\atitle$), and FastText~($\atitle$) performed better. This makes sense because the clickbait problem arises from the fact that the title of the article is ambiguous or misleading \cite{Marquez1980How}. We can also see that the federated learning models perform better. It means that content still has valuable information for the clickbait detection task. It also proved that Clickbait Federated Learning can effectively utilize non-shared data. We can conclude that $PartyA$ and $PartyB$ can cooperate effectively and get a better clickbait detection model no matter what kind of implementation of feature extractor by Clickbait Federated Learning. The best performance in Table \ref{tb3} is FedHHN.

\begin{table}[!ht]
	\caption{The experimental results of the models trained by Clickbait Federated Learning in the Data Island and the models trained in the ideal situation.}\label{tb4}
	\centering
	\begin{tabular}{|c|c|c|}
		\hline
		\multirow{2}{*}{\textbf{Model}} & \multicolumn{2}{c|}{\textbf{Measure}}\\
		\cline{2-3}
		& ROC-AUC & F1 \\
		\hline
		TextCNN($\atitle$\&$\acontent$) & 0.66185 & 0.51734\\
		\hline
		FedCNN & 0.65694 & 0.50881\\
		\hline
		TextRNN($\atitle$\&$\acontent$) & 0.65026 & 0.49115\\
		\hline
		FedRNN & 0.64627 & 0.47744\\
		\hline
		TextSAN($\atitle$\&$\acontent$) & 0.64851 & 0.48759\\
		\hline
		FedSAN & 0.64888 & 0.48603\\
		\hline
		FastText($\atitle$\&$\acontent$) & 0.64489 & 0.47466\\
		\hline
		FedFastText & 0.64224 & 0.46843\\
		\hline
		HHN & \textbf{0.67224} & \textbf{0.53155}\\
		\hline
		FedHHN & \textbf{0.66835} & \textbf{0.52898}\\
		\hline
	\end{tabular}
\end{table}

As shown in Table \ref{tb4}, the ROC-AUC and F1-score of the models trained by Clickbait Federated Learning in the Data Island situation are close to those of the models trained by the traditional training method in the ideal situation, which matches with our analysis using the Chain Rule in section 4.2. This illustrates that Clickbait Federated Learning can effectively utilize non-shared data in the Data Island situation and train a federated learning model which is comparable to the model with the same implementation of the model architecture trained using the traditional training method in the ideal situation. We thus come to the conclusion that Clickbait Federated Learning represents a desirable solution for the Data Island problem.

What is more, the fact that the performance of those models trained by title and content is better than those models trained only with title and those models trained only with content, which is consistent with the belief that the clickbait detection model needs title and content together as input. The fact that FedHHN performs better than TextCNN~($\atitle$\&$\acontent$), TextRNN~($\atitle$\&$\acontent$), TextSAN~($\atitle$\&$\acontent$), and FastText~($\atitle$\&$\acontent$), which is trained in the ideal situation, shows the superiority of Federated Hierarchical Hybrid Networks.

\section{Conclusion}
In this paper, we propose Federated Hierarchical Hybrid Networks for clickbait detection. Federated Hierarchical Hybrid Networks can be considered as Hierarchical Hybrid Networks trained by Clickbait Federated Learning. Hierarchical Hybrid Networks utilize not only the features of the title and content but also the complex connection between the title and content for detecting clickbait. Clickbait Federated Learning can effectively utilize non-shared data in the Data Island setting and train a federated model which is comparable to the model with the same architecture trained using the traditional training method in the ideal situation. It thus represents a desirable solution for the Data Island problem and this method can be extended to any similar multi-input classification tasks with non-shared data. Our experimental results show that Federated Hierarchical Hybrid Networks performed well on clickbait detection tasks.

However, the federated model we get from Clickbait Federated Learning depends on the input from both sides when it makes predictions. How to get a federated model without this dependency is our future work. Moreover, we are also interested in how to apply Clickbait Federated Learning in other multi-input classification tasks with non-shared data.

\bibliography{tacl2018}
\bibliographystyle{acl_natbib}

\end{document}